\documentclass[pdflatex,sn-mathphys-num]{sn-jnl}
\usepackage{graphicx}%
\usepackage{multirow}%
\usepackage{amsmath,amssymb,amsfonts}%
\usepackage{amsthm}%
\usepackage{mathrsfs}%
\usepackage[utf8]{inputenc}
\usepackage[title]{appendix}%
\usepackage{xcolor}%
\usepackage{textcomp}%
\usepackage{manyfoot}%
\usepackage{booktabs}%
\usepackage{algorithm}%
\usepackage{algorithmicx}%
\usepackage{algpseudocode}%
\usepackage{listings}%
\theoremstyle{thmstyleone}%

\usepackage{natbib}
\theoremstyle{thmstyletwo}%
\theoremstyle{thmstylethree}%

\usepackage{cite}
\usepackage{braket}
\usepackage{bm}
\usepackage{comment}

\newcommand{\pto}[1]{\left( #1 \right)}
\newcommand{\pq}[1]{\left[ #1 \right]}

\newcommand{\opc}[2]{\hat{#1}^\dagger_{#2}}
\newcommand{\opd}[2]{\hat{#1}_{#2}}

\begin{document}
\title{Quantum-limited estimation of the frequency shift between two interfering photons by time sampling of their quantum beats}

\author[1]{Luca Maggio}
\author[2,4]{Danilo Triggiani}
\author[3,4]{Paolo Facchi}
\author*[1,5]{Vincenzo Tamma}
\email{vincenzo.tamma@port.ac.uk}
\affil[1]{School of Mathematics and Physics, University of Portsmouth, Portsmouth PO1 3HF, United Kingdom}
\affil[2]{Dipartimento Interateneo di Fisica, Politecnico di Bari, Bari 70126, Italy}
\affil[3]{Dipartimento di Fisica, Universit\`{a} di Bari, I-70126 Bari, Italy}
\affil[4]{INFN, Sezione di Bari, I-70126 Bari, Italy}
\affil[5]{Institute of Cosmology and Gravitation, University of Portsmouth, Portsmouth PO1 3FX, United Kingdom}

\abstract{We present a sensing scheme for estimating the frequency difference of two non-entangled photons. The technique consists of time-resolving sampling measurements at the output of a beam splitter. With this protocol, we can achieve the ultimate precision for the estimation of the frequency shift between two non-entangled photons, overcoming the limits in precision and the range of detection of frequency-resolving detectors employed in standard direct measurements of the frequencies. The sensitivity can be increased by increasing the coherence time of the photons. We show that, already with $\sim 1000$ sampling measurements, the Cram\'{e}r-Rao bound for the specific input state chosen is saturated independently of the value of the difference in frequency.}

\maketitle

\section{Introduction}\label{sec1}
When two photons impinge on a balanced beam splitter from its two input ports, a well-known two-photon interference effect occurs~\citep{PhysRevLett.59.2044, shih1988new,10.1088/1402-4896/adaf7b}. In particular, if the photons are completely identical in all their degrees of freedom, they will always reach one of the two output ports simultaneously (they will always bunch). This phenomenon has been widely applied in quantum information, e.g., quantum computing~\citep{kok2007linear,barz2012demonstration}, quantum key distribution~\citep{tang2014measurement,guan2015experimental}, quantum repeater~\citep{sangouard2011quantum,hofmann2012heralded}, and quantum coherence tomography~\citep{teich2012variations}.

When the two photons are not completely identical, the probability of them reaching two different output ports (the probability of having a coincidence event) is not zero and depends on the overlap between the two single-photon states. Thus, it is possible to retrieve the value of a physical property encoded in such overlap by measuring the coincidence rate. In fact, this interference effect, due to its versatility and robustness against background noise, group velocity dispersion~\citep{PhysRevA.45.6659}, and phase perturbations~\citep{PhysRevA.104.053704}, has several applications in quantum metrology, and it has been used to evaluate the time delays between two non-entangled photons~\citep{lyons2018attosecond}, frequency~\citep{PhysRevA.91.013830,Jin:15,Gianani_2018,fabre2021parameter}, and polarization~\citep{harnchaiwat2020tracking,sgobba2023optimal}.

More recently, it has been shown that it is possible to increase the precision of the estimation of such parameters by encoding them in an entangled state\citep{chen2019hong, meskine2024approaching}. With this technique, the precision increases because the entangled states push the ultimate precision limit by a scaling factor related to the amount of photon in the probe state \citep{PhysRevLett.131.030801}. 

Another line of research aims to maximize the sensitivity in estimating relative photonic parameters, such as time delay or transverse displacement, by resolving their frequencies or transverse momenta~\citep{triggiani2023ultimate, triggiani2024estimation}. The ultimate precision for the estimation of these parameters encoded in a non-entangled-two-photon state is achieved under the assumption of unit detector efficiency and perfect overlap of all the photonic degrees of freedom, but the one to measure.

However, achieving the ultimate quantum precision in the measurement of the photonic frequencies remains an experimental challenge. Indeed, standard techniques that rely on the direct measurement of such frequencies are limited by the precision of the employed frequency-resolved detectors. A first attempt to circumvent direct measurements has been made by using a two-photon interferometric technique in which the temporal delay of the two photons is resolved~\citep{duquennoy2022real,duquennoy2023singular}. The results of the measurements are then analyzed using singular spectrum analysis to establish an asymmetry between the parameters of interest and those that act as nuisance~\citep{suzuki2020nuisance}. However, the authors do not saturate the Cram\'{e}r-Rao bound, and as a consequence they do not achieve the ultimate precision. This problem can be the result of two main factors: First of all, the bunching events are discarded, and only the coincidence events are used for the estimation of the frequency shift between the two photons. Furthermore, the estimator found with the singular spectrum analysis itself might not be adequate to saturate the Cram\'{e}r-Rao bound. 

In this work, we present a two-photon interferometry technique for measuring the frequency shift between two non-entangled photons with which, for the two non-entangled-photon-input state here proposed, the ultimate quantum precision is achieved. The sensing scheme is based on sampling time-resolving measurements of these photons after they impinge on a beam splitter. The precision that we can reach with such a scheme is independent of the values of the relative photonic frequency we wish to estimate. This sensing protocol can have applications in vibrometry~\citep{rehain2021single}, characterization of biological material, or optical coherence tomography~\citep{Kolenderska:20}.

In this Letter, we describe the experimental setup in detail and evaluate its efficiency in a quantum metrology framework. With this setup, we can obtain the ultimate precision for the estimation of the frequency shift between two non-entangled photons under the assumption of unit detector efficiency and perfect overlap of all the degrees of freedom of the two photons but the frequency. Also, we discuss the regime in which this precision is not achieved. We provide a comparison between this setup and a non-resolving protocol, proving that time-resolution measurements improve the estimation, especially in a high frequency-shift regime. Already with $\sim1000$ sampling measurements, the Cram\'{e}r-Rao bound is saturated, independently of the value of the frequency shift.
\begin{figure}
\centering
  \includegraphics[width=100mm]{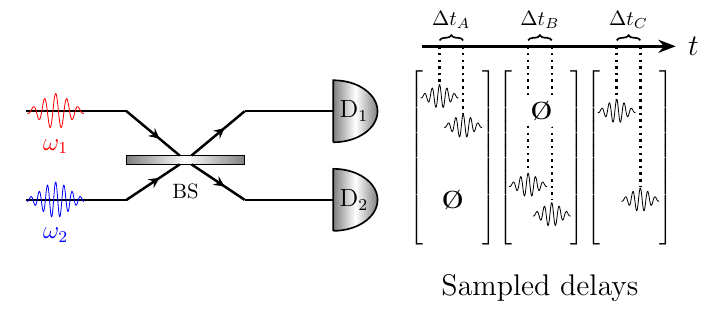}
  \caption{Sensing scheme. Two photons, with central frequencies $\omega_1$ and $\omega_2$, described in the temporal domain in~\ref{eq:input} enter in the input channel of a 50:50 beam splitter (BS). Then, their temporal delay is resolved with two detectors $D_1$ and $D_2$. Here, for example, three repetitions of the experiment are illustated, in which the temporal delays $\Delta t_A$, $\Delta t_B$, $\Delta t_C$, are recorded. These delays, correspond to a bunching event, a coincidence event and a bunching event, respectively.}
  \label{fig:setup}
\end{figure} 

\section{Experimental setup}
\label{SectionII}
In this section, we describe the experimental setup in \figurename~\ref{fig:setup}. A pair of photons impinges on a balanced beam splitter, respectively, through its two input ports. The photons are prepared with a temporal amplitude $\psi_s\pto{t}= \psi\pto{t} \mathrm{e}^{-i\omega_s t}$, where $ \psi\pto{t}=\left\vert \psi\pto{t} \right\vert$ with $s=1,2$. The parameter $\omega_s$ is the center of the s-th photon frequency distribution. The bosonic operators $\opc{a}{s}(t_s)$ and $\opc{b}{s}(t_s)$ are function of the input channel $s$ at time $t_s$ ($s=1,2$). They are defined as the Fourier transforms of the bosonic operators in the frequency domain, $\opc{\tilde{a}}{s}(\Omega)$ and $\opc{\tilde{b}}{s}(\Omega)$, by
\begin{equation}
    \opc{a}{s}(t_s)=\frac{1}{\sqrt{2\pi}}\int_{\mathbb{R}} d\Omega \opc{\tilde{a}}{s}(\Omega)\mathrm{e}^{i\Omega t_s},\qquad \opc{b}{s}(t_s)=\frac{1}{\sqrt{2\pi}}\int_{\mathbb{R}} d\Omega \opc{\tilde{b}}{s}(\Omega)\mathrm{e}^{i\Omega t_s}.
\end{equation}
In this way, for example, the state of the $s$-th photon in the time domain (with creation operator $\opc{a}{s}(t)$) and in the frequency domain (with creation operator $\opc{\tilde{a}}{s}(\Omega)$) are related by the equation
\begin{align}
    \begin{split}
        \ket{\psi_s}&=\int_{\mathbb{R}} dt \psi_s(t)\opc{a}{s}(t)\ket{0}\\
        &=\int_{\mathbb{R}} dt \psi(t)\mathrm{e}^{-i\omega_s t}\frac{1}{\sqrt{2\pi}}\int_{\mathbb{R}} d\Omega \opc{\tilde{a}}{s}(\Omega)\mathrm{e}^{i\Omega t}\ket{0}\\
        &=\int_{\mathbb{R}} d\Omega\left(\frac{1}{\sqrt{2\pi}}\int_{\mathbb{R}} dt \psi(t)\mathrm{e}^{i(\Omega-\omega_s) t}\right)\opc{\tilde{a}}{s}(\Omega)\ket{0}\\
        &=\int_{\mathbb{R}} d\Omega\tilde{\psi}(\Omega-\omega_s)\opc{\tilde{a}}{s}(\Omega)\ket{0},
    \end{split}
\end{align}
Where $\tilde{\psi}(\Omega)=\frac{1}{\sqrt{2\pi}}\int_{\mathbb{R}} dt \psi(t)\mathrm{e}^{i\Omega t}$ is the Fourier transform of $\psi(t)$.
The commutation relation between $\opd{a}{s}(t_1)$ and $\opc{b}{s'}(t_2)$ is $\pq{\opd{a}{s}(t_1),\opc{b}{s'}(t_2)}=\sqrt{\nu}\delta_{ss'}\delta\pto{t_1-t_2}$, where $\delta_{ss'}$ and $\delta\pto{t_1-t_2}$ are, respectively, the Kroenecker delta as a function of the input channels and the Dirac delta in the time domain. The parameter $\nu\in\pq{0,1}$ represents the degree of indistinguishability of the two photons in any other degree of freedom, except time. Analogously, in the frequency domain, $\pq{\opd{\tilde{a}}{s}(\Omega_1),\opc{\tilde{b}}{s'}(\Omega_2)}=\sqrt{\nu}\delta_{ss'}\delta\pto{\Omega_1-\Omega_2}$. The two-photon input state reads
\begin{figure}
\centering
 \includegraphics[width=100mm]{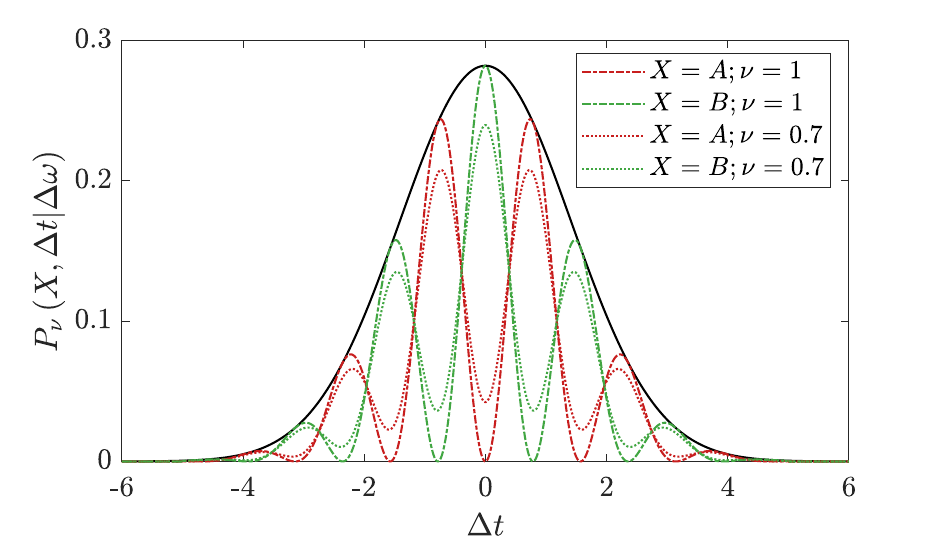}
 \caption{Plot of the probability distribution in eq.~\ref{eq:probs}. The temporal distribution $\left\vert\psi\pto{t}\right\vert^2$ is considered Gaussian with unitary variance. This yields a Gaussian envelope $C\pto{\Delta t}$, represented by the black line. The choice of the variance fixed a natural scale for $\Delta t$ and for $\Delta\omega=4/\tau$}.
\label{fig:probs}
\end{figure}

\begin{equation}
 \ket{\bm{\psi}}=\int_{\mathbb{R}^2} dt_1dt_2 \psi_1\pto{t_1}\psi_2\pto{t_2}\opc{a}{1}(t_1)\opc{b}{2}(t_2)\ket{0},\label{eq:input}
\end{equation}
where $\ket{0}=\ket{0}_1\otimes\ket{0}_2$ is the vacuum state.

Two detectors are connected to the two output channels of the beam splitter. They are sensitive to the time of arrival of the two photons. This experimental setup does not directly measure the frequency of the photons; instead, the sensing protocol relies on time-sampling measurements~\citep{legero2003time}. The measure of the time delay between the two photons can be done with a precision of the order of the ps~\citep{korzh2020demonstration,instruments2040019,esmaeil2017single,wu2017improving}. The only requirement for the precision in time $\delta t$ is to be high enough to erase the indistinguishability in time of the two photons, e.g. for Gaussian distribution in time with variance $\tau$, the requirement for the estimation of $\Delta\omega=\omega_2-\omega_1$ is
\begin{equation}
\delta t\ll \frac{1}{|\Delta\omega|}\qquad \mathrm{and} \qquad \delta t\ll \tau.\label{main:cond}
\end{equation}
For example, this means that already by employing detectors with a time resolution $\delta t \approx10$ ps, it is possible to estimate a range of values of $\Delta\omega$ up to the range $10$ GHz, or up to the range $100$ GHz without changing the experimental setup if one can employ detectors with resolution $\delta t$ of the order of 1 ps.  In fact, state-of-the-art time-resolving detectors can already achieve a resolution of a few picoseconds~\citep{wu2017improving,instruments2040019,esmaeil2017single,korzh2020demonstration}. As long as this bound in $\Delta\omega$ is respected, the values of $\omega_1$ and $\omega_2$ can be in any frequency range, overcoming the range limitations of standard spectrometers~\citep{Davis:17,PhysRevApplied.14.014052}.
Furthermore, the second condition in Eq.\eqref{main:cond}
is easily satisfied with the state-of-the-art single-photon sources~\citep{Zhao:14,10.1063/5.0217815}. 

In this setup, the two photons can arrive on the same output channel (bunching, labeled with the letter A), or they can arrive on two different output channels (coincidence, labeled with the letter B). In addition, the two photons impinge on the detectors with a certain time delay $\Delta t$. The probability of having $X=A,B$ and a specific $\Delta t$ is
\begin{equation}
 P_\nu\pto{X,\Delta t|\Delta\omega}=\frac{C\pto{\Delta t}}{2}\Bigg(1+\nu\alpha\pto{X}\cos\pto{\Delta\omega\Delta t}\Bigg),\label{eq:probs}
\end{equation}
where $\alpha\pto{A}=1$, $\alpha\pto{B}=-1$, and $C\pto{\Delta t}$ is the beats envelope, whose shape can be evaluated by using $\psi\pto{t}$. We plot this probability as a function of $\Delta t$ in \figurename~\ref{fig:probs}. In particular, for two Gaussian distributions in time with variance $\tau$, $C\pto{\Delta t}$ assumes the form
\begin{equation}
 C\pto{\Delta t}=\frac{1}{\sqrt{4\pi\tau^2}}\mathrm{exp}\pq{-\frac{\Delta t^2}{4\tau^2}}.\label{eq:gaussenv}
\end{equation}

By introducing the efficiency of the detectors $\gamma\in\pq{0,1}$ (assumed to be equal for both detectors), it is possible to define the probability of measuring $0,1,2$ photons in output with probabilities that are, respectively

\begin{align}
\begin{split}
&P_0=\pto{1-\gamma}^2\\
&P_1=2\gamma\pto{1-\gamma}\\
&P_2\pto{X,\Delta t|\Delta\omega}=\gamma^2P_\nu\pto{X,\Delta t|\Delta\omega}, \qquad X=A,B
.\label{eq:gammaprob}
\end{split}
\end{align}

Only the probability $P_2\pto{X,\Delta t|\Delta\omega}$ is function of the parameter $\Delta\omega$. This probability is proportional to the one previously found in Eq.~\eqref{eq:probs} by a factor $\gamma^2$.
\section{Ultimate quantum sensitivity}
An unbiased estimator of the frequency shift $\widetilde{\Delta\omega}$ is affected by an error described by the variance $\mathrm{Var}[\widetilde{\Delta\omega}]$. This variance will always be bounded below by the Cram\'{e}r-Rao bound, which represents the maximum precision achievable by the sensing protocol and it is a function of the Fisher information $F_\nu\pto{\Delta\omega}$ and the number of iterations of the experiment $N$~\citep{cramer1999mathematical, rohatgi2015introduction}. This bound is always saturated in the asymptotic limit of large $N$. Ultimately, the Cram\'{e}r-Rao bound is bounded from below by the quantum Cram\'{e}r-Rao bound, which is the maximum precision achievable by any scheme, independently of the chosen type of measurement, and it depends on the quantum Fisher information $H\pto{\Delta\omega}$ and $N$~\citep{helstrom1969quantum,holevo2011probabilistic}. Therefore, the following inequalities hold
\begin{equation}
\mathrm{Var}\pq{\widetilde{\Delta\omega}}\geq \frac{1}{N F_\nu\pto{\Delta\omega}}\geq \frac{1}{N H\pto{\Delta\omega}}. \label{eq:bounds}
\end{equation}
In particular, by using the probabilities in Eq.~\ref{eq:gammaprob}, the Fisher information assumes the form
\begin{equation}
F_{\nu=1}\pto{\Delta\omega}=\gamma^2H\pto{\Delta\omega}=2\gamma^2\tau^2\label{eq:QCRB},
\end{equation}
proving that under the condition of unit detection efficiency, this sensing protocol reaches the ultimate precision for $\nu=1$. This happens because for $\nu=1$ the detectors cannot distinguish at all the two input photons, enhancing the two-photon interference. Interestingly, the overlap of the density distributions of the two photons in the frequency domain does not affect the asymptotic sensitivity of the sensing scheme. Also, the dependence of the Fisher information only on the coherence time is an advantage with respect to the standard direct measurements, since these operate only in a specific range of frequencies and their precision is limited by their finite resolution. Remarkably, our technique allows us to obtain a precision of already the order of $0.1$ MHz (which corresponds, for example, in the visible range to a precision of the order of 20 am for wavelengths of the order of 800 nm) for $\tau^2N=10^7 \mathrm{ns}^2$, achievable, for example, with a coherence time $\tau=100$ ns~\citep{mckeever2004deterministic,wilk2007polarization,keller2004continuous} and just a number of sampling measurements $N\sim 10^3$, or increasing accordingly the number of measurements if photons of smaller coherence times are employed. Therefore, this technique outperforms the precision of most standard spectrometers in the state-of-the-art~\citep{Davis:17,PhysRevApplied.14.014052}. 
\section{Contribution to the sampled time delay}
\label{SectionIII}
\begin{figure}
\centering
\includegraphics[width=100mm]{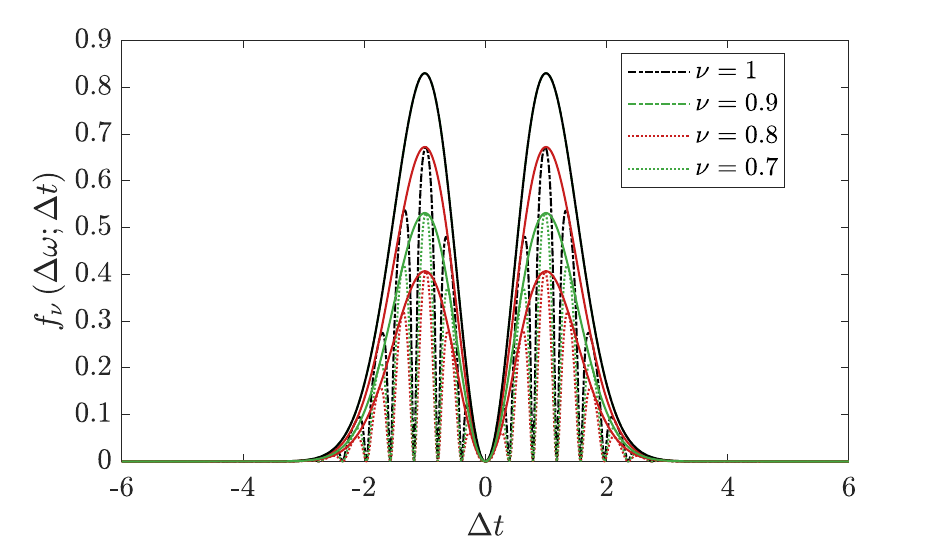}
\caption{Contributions $f_\nu\pto{\Delta\omega ; \Delta t}$ in the Fisher information as a function of the sampled time delay $\Delta t$ for different values of the parameter $\nu$. The plot is for a Gaussian distribution in time of the two photons $\left\vert \psi\pto{t}\right\vert^2$ with variance $\tau$. The variance fixes a natural scale for $\Delta t$ and for $\Delta\omega=4/\tau_{\Delta t}$. The solid lines represent the envelope of $f_\nu\pto{\Delta\omega; \Delta t}$ and show the contribution of the Fisher information that is significant or negligible.}
\label{fig:cont}
\end{figure}
Our sensing protocol remains effective even by using as input two photons that are partially distinguishable in all the degrees of freedom but the time. This is the case in which $\nu<1$ and the Fisher information reads
\begin{align}
\begin{split}
F_{\nu}\pto{\Delta\omega}&=\gamma^2\int d t f_\nu\pto{\Delta\omega ;  t}\\
&=\gamma^2\int_{\mathbb{R}}d t C\pto{t}t^2\beta_\nu\pto{\Delta\omega t},\label{eq:fi}
\end{split}
\end{align}
where $f_\nu\pto{\Delta\omega ; \Delta t}$ is the contribution of each $\Delta t$ to the Fisher information. Also, the function $\beta_\nu\pto{\xi}$ is defined as follows (for more details see Appendix~\ref{app:Fi}):
\begin{equation}
\beta_\nu\pto{\xi}=\frac{\nu^2\sin^2\pto{\xi}}{1-\nu^2\cos^2\pto{\xi}}.
\end{equation}
For Gaussian distributions, the Fisher information~\eqref{eq:fi} specializes into
\begin{equation}
F^G_\nu\pto{\Delta\omega}=\frac{2F_{\nu=1}\pto{\Delta\omega}}{\sqrt{\pi}}\int_{\mathbb{R}}\mathrm{d}\kappa\, \mathrm{e}^{-\kappa^2}\kappa^2\beta_\nu\pto{2\tau\Delta\omega\kappa}.\label{eq:fig}
\end{equation}
We plot the contribution of the Fisher information $f_\nu\pto{\Delta\omega ; \Delta t}$ in \figurename~\ref{fig:cont}, where we show the amount of information obtained for each sampled time delay. In particular, we show that $f_\nu\pto{\Delta\omega ; \Delta t}$ is concentrated within its envelope, that is, $\nu^2 C\pto{\Delta t}\Delta t^2/2\tau^2$. The envelope is independent of $\Delta\omega$, which means that the metrological scheme does not need to be calibrated according to the parameter to estimate $\Delta\omega$.

\begin{figure}
\centering
\includegraphics[width=100mm]{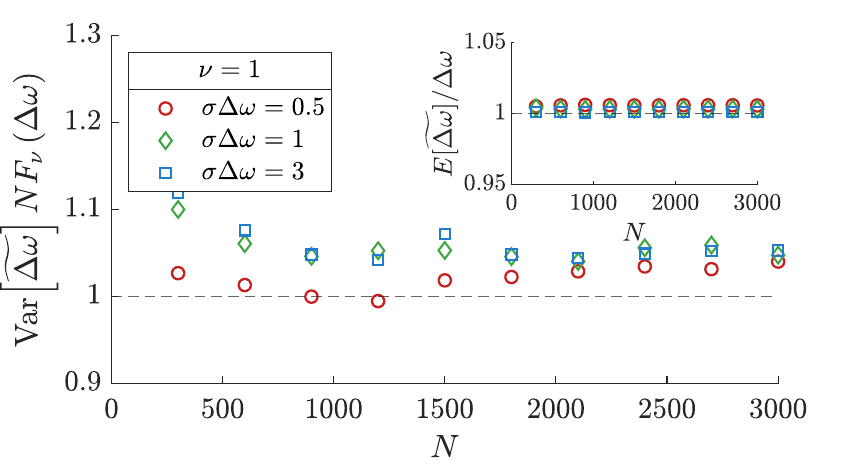}
  \includegraphics[width=100mm]{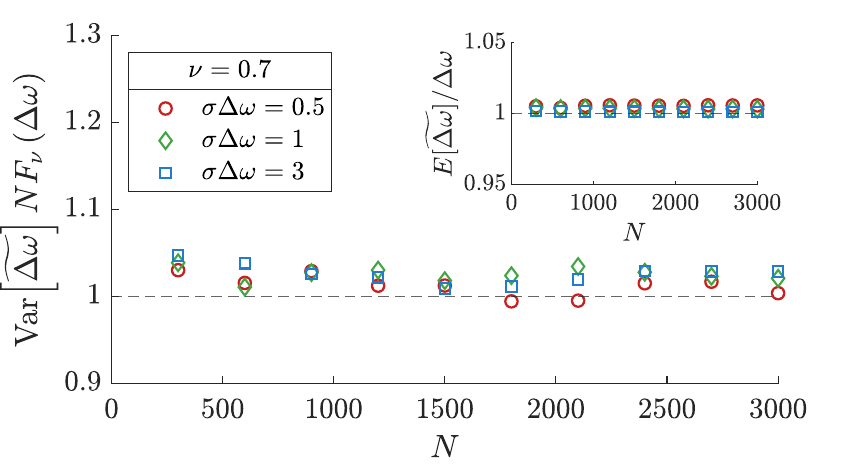}
  \caption{Simulation of the variance normalized with respect to the Cram\'{e}r-Rao bound for $\nu=1$ (top figure), and $\nu=0.7$ (bottom figure), for values of $\sigma\Delta\omega=0.5,1,3$, respectively represented with a circle, a rhombus and a square marker. In the inset it is shown the simulation of the Expected value of $\Delta\omega$ normalized. The simulation has been executed by repeating for each point the measure process $10^4$ times. It is possible to show that even for an amount of data $N\sim 10^3$ the Cram\'{e}r-Rao bound is saturated, and the bias is inferior to the $1\%$.}
  \label{fig:sim}
\end{figure}

We show the efficiency of the sensing scheme in \figurename~\ref{fig:sim} which exploits the simulated variance of the likelihood estimator~\citep{rohatgi2015introduction} normalized to the Cram\'{e}r-Rao bound. It is evident that for a number of sampling measurements $N\sim 1000$ the Cram\'{e}r-Rao bound is saturated. In the insets, we show the expected value $E\pq{\widetilde{\Delta\omega}}$ of $\Delta\omega$, proving that the bias is inferior to $1\%$ of the value of $\Delta\omega$ in the regime of $N\sim 1000$. This result is independent of the values of $\Delta\omega$ and $\nu$.

\section{Comparison with the non-resolving interferometric scheme}\label{SectionIV}
\begin{figure}
\centering
  \includegraphics[width=100mm]{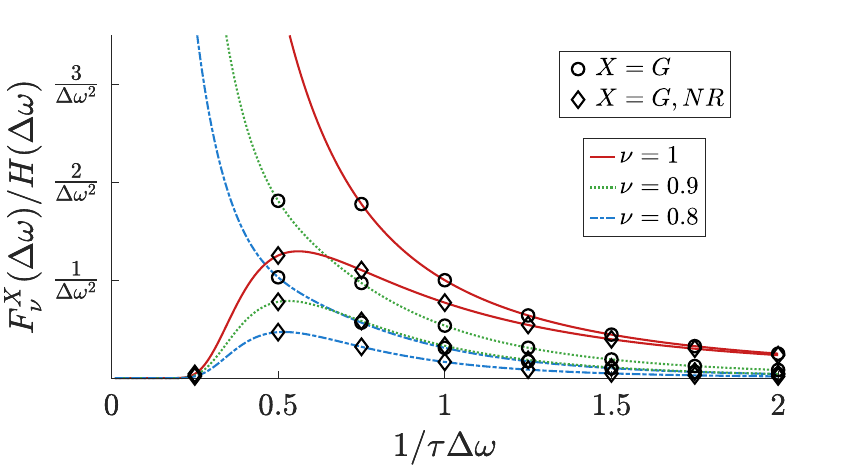}
  \caption{Plot of Fisher information $F_\nu^G\pto{\Delta\omega}$ in Eq.~\eqref{eq:fig} and $F_\nu^{G,NR}\pto{\Delta\omega}$ in Eq.~\eqref{eq:fignr} vs $1/\Delta\omega$, for different values of the parameter $\nu=1,0.9,0.8$. The non resolving Fisher information reaches the Quantum Cram\'{e}r-Rao bound at $\nu=1$ and for $\Delta\omega\ll 1/\tau$. Instead, for $\Delta\omega\gg 1/\tau$, the non-resolving approach cannot retrieve any information of the parameter $\Delta\omega$. }
  \label{fig:QCRB}
\end{figure}

In this section, we compare the effectiveness of this protocol with the one of a sensing scheme that does not resolve the time delay of the photons, as in~\citep{fabre2021parameter}. In case of a Gaussian distribution in time for both photons, the Fisher information for non-resolving time delay measurements and unit detector precision is
\begin{equation}
F^{G,NR}_\nu\pto{\Delta\omega}=\frac{4\nu^2\tau^4\Delta\omega^2}{\mathrm{e}^{2\tau^2\Delta\omega^2}-\nu^2}=H\pto{\Delta\omega}\frac{2\nu^2\tau^2\Delta\omega^2}{\mathrm{e}^{2\tau^2\Delta\omega^2}-\nu^2}.\label{eq:fignr}
\end{equation}
Both $F^G_\nu\pto{\Delta\omega}$ and $F^{G,NR}_\nu\pto{\Delta\omega}$ are represented in \figurename~\ref{fig:QCRB} as a function of $1/\tau\Delta\omega$ for different values of $\nu$ and for $\gamma=1$. It is possible to show that, due to the convexity of the Fisher information, $F^G_\nu\pto{\Delta\omega}\geq F^{G,NR}_\nu\pto{\Delta\omega}$. For $\tau\Delta\omega\ll1$, $F^{G,NR}_\nu\pto{\Delta\omega}$ approaches $F^G_\nu\pto{\Delta\omega}$ and in particular, for $\nu=1$, the sensitivities of the resolving and non-resolving schemes are equal. 

In the case of $\tau\Delta\omega\gg 1$, i.e., the case in which the photons do not overlap in their frequency distribution, the Fisher information of the resolving sensing scheme is proportional to the quantum Fisher information by a factor that depends only on $\nu$ and $\gamma$. In this case, as shown in Appendix~\ref{app:Fi},
\begin{equation}
F_\nu\pto{\Delta\omega}=\gamma^2\pto{1-\sqrt{1-\nu^2}}F_{\nu=1}\pto{\Delta\omega}.
\end{equation}
Instead, the Fisher information of the non-resolving sensing scheme cannot retrieve any information about $\Delta\omega$ for $\tau\Delta\omega\gg 1$.

This happens because the non-resolving scheme requires the photons to be as similar as possible in order to maximize the interference effect. Therefore, the non-resolving case is highly sensitive only for small values of the frequency shift. On the other hand, the indistinguishability at the detector is preserved for any frequency shift when resolving the times. 

\section{Conclusions}
\label{SectionV}
We demonstrate the efficiency of an interferometric scheme based on time-resolving sampling measurements for the estimation of the frequency shift $\Delta\omega$ of two photons. We show that in the regime of $N\sim 1000$ the Cram\'{e}r-Rao bound is approximately saturated and the bias is inferior to $1\%$ of the value of $\Delta\omega$, independently of the overlap between the photons and $\Delta\omega$ itself.

In fact, the efficiency of the detection does not depend on $\Delta\omega$. This is more evident when the sensing scheme is compared with a non-resolving interferometric technique, which is not efficient for large values of $\Delta\omega$. Instead, the precision of the estimation depends on the coherence time of the photons. That makes the precision virtually without an upper limit, especially considering techniques that can increase the coherence time of the photons.

Furthermore, this technique is not constrained by the detector frequency resolution required to directly measure the difference in frequency between the two photons. Instead, it varies on the efficiency of the detector in time. However, this requirement is considered only for imposing the photons indistinguishability at the detector. Having only this constraint in the detection allows us to outperform the standard spectrometers on a broader range. 

\hypersetup{bookmarksdepth=3}
\bmhead{Acknowledgements}
This project is partially supported by Xairos Systems Inc. VT also acknowledges partial support from the Air Force Office of Scientific Research under award number FA8655-23-17046. PF was partially supported by Istituto Nazionale di Fisica Nucleare (INFN) through the project ``QUANTUM'', by the Italian National Group of Mathematical Physics (GNFM-INdAM), and by the Italian funding within the ``Budget MUR - Dipartimenti di Eccellenza 2023--2027'' - Quantum Sensing and Modelling for One-Health (QuaSiModO). DT acknowledges the Italian Space Agency (ASI, Agenzia Spaziale Italiana) through the project Subdiffraction Quantum Imaging (SQI) n. 2023-13-HH.0.

\bmhead{Data availability statement}
Supporting research data are available on reasonable request from the corresponding author V.T.

\begin{appendices}
 \setcounter{equation}{0} 
\section{Evaluation of the probabilities in Eq.~\eqref{eq:probs}}
\subsection{Evaluation of the correlation function}
In this section we evaluate the probabilities in Eq.~\eqref{eq:probs} of having bunching or coincidence in the sensing scheme in \figurename~\ref{fig:setup} by using the correlation function  
\begin{equation}
G^{\pto{2}}_{C_1C_2}\pto{\alpha_1,t_1;\alpha_2,t_2}=\bra{\bm{\psi}}\opd{E}{C_1}^{-\pto{\alpha_1}}(t_1)\opd{E}{C_2}^{-\pto{\alpha_2}}(t_2)\opd{E}{C_1}^{+\pto{\alpha_1}}(t_1)\opd{E}{C_2}^{+\pto{\alpha_2}}(t_2)\ket{\bm{\psi}}.\label{app:g2}
\end{equation}
Here, the state $\ket{\bm{\psi}}$ correspond to the input state in Eq.~\eqref{eq:input}, that we rewrite below for clarity:
\begin{equation}
\ket{\bm{\psi}}=\int_{\mathbb{R}^2} dt'_1dt'_2 \psi_1\pto{t'_1}\psi_2\pto{t'_2}\opc{a}{1}(t'_1)\opc{b}{2}(t'_2)\ket{0},\label{app:input}
\end{equation}
with $\psi_s\pto{t}=\left\vert \psi\pto{t} \right\vert \mathrm{e}^{-i\omega_s t}$. Also, $C_1,C_2$ are respectively the first and the second output of the beam splitter; $t_1$ and $t_2$ are the time of detection of each photon; $\opd{\alpha}{s}$, with $s=1,2$ represent the orthogonal modes in which the two photons can be, i.e. $\opd{a}{s}, \opd{d}s$, such that $\opd{b}{s}=\sqrt{\nu}\opd{a}{s}+\sqrt{1-\nu}\opd{d}{s}$. We write the field operators in the Heisemberg picture the following way
\begin{align}
\begin{split}
&\opd{E}{C_1}^{+\pto{\alpha}}(t_1)=\sum_{s=1,2}\frac{\mathrm{e}^{i\Phi\pto{s,C_1}}}{\sqrt{2}}\opd{\alpha}{s}(t_1)=\sum_{s=1,2}\opd{E}{sC_1}^{+\pto{\alpha}}(t_1) , \\&\opd{E}{C_2}^{+\pto{\alpha}}(t_1)=\sum_{s=1,2}\frac{\mathrm{e}^{i\Phi\pto{s,C_2}}}{\sqrt{2}}\opd{\alpha}{s}(t_1)=\sum_{s=1,2}\opd{E}{sC_2}^{+\pto{\alpha}}(t_1) , \\&\hat{\alpha}=\hat{a},\hat{d}; \pto{\opd{E}{C_1}^{+\pto{\alpha}}(t_1)}^\dagger=\opd{E}{C_1}^{-\pto{\alpha}}(t_1), \pto{\opd{E}{C_2}^{+\pto{\alpha}}(t_1)}^\dagger=\opd{E}{C_2}^{-\pto{\alpha}}(t_1)\label{app:fieldop}
\end{split}
\end{align}
where $s$ represents the photon in the input channel $s$ and $\Phi$ is the phase acquired to the beam splitter.

Since $\ket{\bm{\psi}}$ is a two-photon state and $\opd{E}{C_1}^{+\pto{\alpha_1}}(t_1)\opd{E}{C_2}^{+\pto{\alpha_2}}(t_2)$ annihilates two photons, $\opd{E}{C_1}^{+\pto{\alpha_1}}(t_1)\opd{E}{C_2}^{+\pto{\alpha_2}}(t_2)\ket{\bm{\psi}}\propto \ket{0}$. Similarly, $\bra{\bm{\psi}}\opd{E}{C_1}^{-\pto{\alpha_1}}(t_1)\opd{E}{C_2}^{-\pto{\alpha_2}}(t_2)\propto\bra{0}$. So, by expanding Eq.~\eqref{app:g2} by using Eq.~\eqref{app:fieldop} we can write
\begin{align}
\begin{split}
G^{\pto{2}}_{C_1C_2}\pto{\alpha_1,t_1;\alpha_2,t_2}&=\sum_{s_i=1,2}\bra{\bm{\psi}}\opd{E}{s_1C_1}^{-\pto{\alpha_1}}(t_1)\opd{E}{s_2C_2}^{-\pto{\alpha_2}}(t_2)\ket{0}\bra{0}\opd{E}{s_3C_1}^{+\pto{\alpha_1}}(t_1)\opd{E}{s_4C_2}^{+\pto{\alpha_2}}(t_2)\ket{\bm{\psi}}\\
&=\sum_{s_i=1,2}G^{\pto{2}}_{s_1s_2s_3s_4C_1C_2}.
\end{split}
\end{align}
By considering that the two input photons are in two different channels, this equation allows us to show that
\begin{align}
G^{\pto{2}}_{11s_3s_4C_1C_2}=G^{\pto{2}}_{22s_3s_4C_1C_2}=G^{\pto{2}}_{s_1s_211C_1C_2}=G^{\pto{2}}_{s_1s_222C_1C_2}=0,
\end{align}
therefore,
\begin{align}
G^{\pto{2}}_{C_1C_2}\pto{\alpha_1,t_1;\alpha_2,t_2}&=\sum_{s_1,s_2=1,2}G^{\pto{2}}_{s_1\sigma\pto{s_1}s_2\sigma\pto{s_2}C_1C_2}\, , \\\sigma\pto{1}=2,\sigma\pto{2}=1.
\end{align}
We can rewrite the four remaining terms of the sum in a simpler form, that is
\begin{align}
G^{\pto{2}}_{C_1C_2}\pto{\alpha_1,t_1;\alpha_2,t_2}&=\left\vert\sum_{s_1=1,2}\bra{0}\opd{E}{s_1C_1}^{+\pto{\alpha_1}}(t_1)\opd{E}{\sigma\pto{s_1} C_2}^{+\pto{\alpha_2}}(t_2)\ket{\bm{\psi}}\right\vert^2.\label{app:g2pt2}
\end{align}
Each term of the sum can be expanded by using Eq.~\eqref{app:input} and Eq.~\eqref{app:fieldop}. We have
\begin{align}\begin{split}
\bra{0}\opd{E}{1 C_1}^{+\pto{a}}(t_1)\opd{E}{2 C_2}^{+\pto{d}}(t_2)\ket{\bm{\psi}}&=\frac{1}{2}\int_{\mathbb{R}^2}dt'_1dt'_2 \mathrm{e}^{i\Phi\pto{1,C_1}}\mathrm{e}^{i\Phi\pto{2,C_2}}\psi_1\pto{t'_1}\psi_2\pto{t'_2}\times\\
&\times\bra{0}\opd{a}{1}(t_1)\opd{d}{2}(t_2)\opc{a}{1}(t'_1)\opc{b}{2}(t'_2)\ket{0}\\
&=\frac{\sqrt{1-\nu}}{2}\mathrm{e}^{i\Phi\pto{1,C_1}}\mathrm{e}^{i\Phi\pto{2,C_2}}\psi_1\pto{t_1}\psi_2\pto{t_2},
\end{split}\end{align}
\begin{align}\begin{split}
\bra{0}\opd{E}{2 C_1}^{+\pto{d}}(t_1)\opd{E}{1 C_2}^{+\pto{a}}(t_2)\ket{\bm{\psi}}&=\frac{1}{2}\int_{\mathbb{R}^2}dt'_1dt'_2 \mathrm{e}^{i\Phi\pto{2,C_1}}\mathrm{e}^{i\Phi\pto{1,C_2}}\psi_1\pto{t'_1}\psi_2\pto{t'_2}\times\\
&\times\bra{0}\opd{d}{2}(t_1)\opd{a}{1}(t_2)\opc{a}{1}(t'_1)\opc{b}{2}(t'_2)\ket{0}\\
&=\frac{\sqrt{1-\nu}}{2}\mathrm{e}^{i\Phi\pto{1,C_1}}\mathrm{e}^{i\Phi\pto{2,C_2}}\psi_1\pto{t_2}\psi_2\pto{t_1},
\end{split}\end{align}
\begin{align}\begin{split}
\bra{0}\opd{E}{s_1C_1}^{+\pto{a}}(t_1)\opd{E}{\sigma\pto{s_1} C_2}^{+\pto{a}}(t_2)\ket{\bm{\psi}}&=\frac{1}{2}\int_{\mathbb{R}^2}dt'_1dt'_2 \mathrm{e}^{i\Phi\pto{s_1,C_1}}\mathrm{e}^{i\Phi\pto{\sigma\pto{s_1},C_2}}\psi_1\pto{t'_1}\psi_2\pto{t'_2}\times\\
&\times\bra{0}\opd{d}{s_1}(t_1)\opd{a}{\sigma\pto{s_1}}(t_2)\opc{a}{1}(t'_1)\opc{b}{2}(t'_2)\ket{0}\\
&=\frac{\sqrt{\nu}}{2}\Big(\delta_{1s_1}\mathrm{e}^{i\Phi\pto{1,C_1}}\mathrm{e}^{i\Phi\pto{2,C_2}}\psi_1\pto{t_1}\psi_2\pto{t_2}+\\
&+\delta_{2s_1}\mathrm{e}^{i\Phi\pto{2,C_1}}\mathrm{e}^{i\Phi\pto{1,C_2}}\psi_1\pto{t_2}\psi_2\pto{t_1}\Big).
\end{split}\end{align}
In addition, recalling that mode $\opc{a}{}$ is orthogonal to mode $\opc{d}{}$ and that $\opc{b}{}=\sqrt{\nu}\opc{a}{}+\sqrt{1-\nu}\opc{d}{}$, we can find that: 
\begin{align}\begin{split}
&\bra{0}\opd{E}{s_1C_1}^{+\pto{d}}(t_1)\opd{E}{\sigma\pto{s_1} C_2}^{+\pto{d}}(t_2)\ket{\bm{\psi}}=0,\\
&\bra{0}\opd{E}{2 C_1}^{+\pto{a}}(t_1)\opd{E}{1 C_2}^{+\pto{d}}(t_2)\ket{\bm{\psi}}=0,\\
&\bra{0}\opd{E}{1 C_1}^{+\pto{d}}(t_1)\opd{E}{2 C_2}^{+\pto{a}}(t_2)\ket{\bm{\psi}}=0,
\end{split}\end{align}

Using these results in Eq.~\eqref{app:g2pt2}, all the second-order correlation functions are
\begin{align}\begin{split}
G^{\pto{2}}_{C_1C_2}\pto{d,t_1;d,t_2}&=0,\\
G^{\pto{2}}_{C_1C_2}\pto{a,t_1;d,t_2}&=G^{\pto{2}}_{C_1C_2}\pto{d,t_1;a,t_2}=\frac{1-\nu}{4}\left\vert\psi\pto{t_2}\psi\pto{t_1}\right\vert^2,\\
G^{\pto{2}}_{C_1C_2}\pto{a,t_1;a,t_2}&=\frac{\nu}{4}\left\vert\psi\pto{t_2}\psi\pto{t_1}\right\vert^2\Big\vert \mathrm{e}^{i\Phi\pto{1,C_1}+i\Phi\pto{2,C_2}}\mathrm{e}^{-i\omega_1 t_1-i\omega_2 t_2}\\&+\mathrm{e}^{i\Phi\pto{2,C_1}+i\Phi\pto{1,C_2}}\mathrm{e}^{-i\omega_1 t_2-i\omega_2 t_1}  \Big\vert^2\\
&=\frac{\nu}{2}\left\vert\psi\pto{t_2}\psi\pto{t_1}\right\vert^2\times\\&\times\pto{1+\mathrm{Re\pto{ \mathrm{e}^{i\pq{\Phi\pto{1,C_1}+\Phi\pto{2,C_2}-\Phi\pto{2,C_1}-\Phi\pto{1,C_2}}}\mathrm{e}^{i\Delta\omega\Delta t}}}}\\
&=\frac{\nu}{2}\left\vert\psi\pto{t_2}\psi\pto{t_1}\right\vert^2\pto{1+\pto{2\delta_{C_1C_2}-1}\cos\Delta\omega\Delta t},
\end{split}\end{align}
where $\Phi\pto{1,C_1}+\Phi\pto{2,C_2}-\Phi\pto{2,C_1}-\Phi\pto{1,C_2}=\pm \pi$ in case of coincidence event ($C_1\neq C_2$) and it is equal to 0 in case of a bunching event ($C_1=C_2$).
\subsection{Evaluation of the probabilities}
After evaluating the second-order correlation functions we can retrieve the probabilities of coincidence (labelled with $A$) and bunching (labelled with $B$) as follows:
\begin{align}\begin{split}
P_\nu\pto{t_1,t_2,A|\Delta\omega}&=\sum_{\alpha_1,\alpha_2=a,d}G^{\pto{2}}_{12}\pto{\alpha_1,t_1;\alpha_2,t_2}\\&=\frac{1}{2}\left\vert\psi\pto{t_2}\psi\pto{t_1}\right\vert^2\pto{1-\nu\cos\Delta\omega\Delta t},\\
P_\nu\pto{t_1,t_2,B|\Delta\omega}&=\frac{1}{2}\sum_{\alpha_1,\alpha_2=a,d}G^{\pto{2}}_{11}\pto{\alpha_1,t_1;\alpha_2,t_2}+G^{\pto{2}}_{22}\pto{\alpha_1,t_1;\alpha_2,t_2}\\&=\frac{1}{2}\left\vert\psi\pto{t_2}\psi\pto{t_1}\right\vert^2\pto{1+\nu\cos\Delta\omega\Delta t},\label{app:probt1t2}
\end{split}\end{align}
where the factor $1/2$ is used for taking into account the symmetry $t_2\leftrightarrow t_1$. Since the sensing protocol is designed for detecting only the time delay, we can define $\tau_{M}=(t_1+t_2)/2$ and $\Delta t=t_2-t_1$, and we can integrate the probabilities over $\tau_{M}$, obtaining
\begin{equation}
    P_\nu\pto{\Delta t,X|\Delta\omega}=\int_{\mathbb{R}}d\tau_{M}P_\nu \pto{\tau_{M}-\frac{\Delta t}{2},\tau_{M}+\frac{\Delta t}{2},X},\,X=A,B.
\end{equation}
By solving the integral, the probabilities in Eq.~\eqref{eq:probs} are,
\begin{equation}
        P_\nu\pto{\Delta t,X|\Delta\omega}=\frac{C\pto{\Delta t}}{2}\pto{1+\nu\alpha\pto{X}\cos\pto{\Delta\omega\Delta t}},\, X=A,B,\label{app:probs}
\end{equation}
where $C\pto{\Delta t}$ is the envelope defined as follows:
\begin{equation}
   C\pto{\Delta t}= \int_{\mathbb{R}}d\tau_{M}\left\vert\psi\pto{\tau_{M}-\frac{\Delta t}{2}}\psi\pto{\tau_{M}+\frac{\Delta t}{2}}\right\vert^2.\label{app:env}
\end{equation}
If we take into account the efficiency of the detectors $\gamma\in\pq{0,1}$ (which is the same for both the detectors), it is possible to register events where 0 photons or only one photon is detected. By defining these probabilities respectively as $P_0$ and $P_1$ and by using Eq.~\eqref{app:probt1t2}, we can write
\begin{align}\begin{split}
&P_0=\pto{1-\gamma}^2\\
&P_1\pto{t_1}=\gamma\pto{1-\gamma}\int_{\mathbb{R}}dt_2 \pto{P_\nu\pto{t_1,t_2,A}+P_\nu\pto{t_1,t_2,B}}\\
&P\pto{t_1,t_2,X}=\gamma^2P_\nu\pto{t_1,t_2,X}, \, X=A,B.
\end{split}\end{align}
Therefore, integrating with respect to $\tau_{M}$, the probabilities become
\begin{align}\begin{split}
&P_0=\pto{1-\gamma}^2\\
&P_1\pto{\Delta t}=2\gamma\pto{1-\gamma}\int d\tau_{M}\left\vert\psi\pto{t_1}\right\vert^2=2\gamma\pto{1-\gamma}\\
&P\pto{\Delta t.X|\Delta\omega}=\gamma^2P_\nu\pto{\Delta t.X|\Delta\omega}, \, X=A,B.
\end{split}\end{align}
Here, the one-photon event and the zero-photon event have a probability that depends only to the efficiency of the detectors. The two-photon event has a probability that is proportional to the square of the efficiency.

 \setcounter{equation}{0} 
\section{Evaluation of the quantum Fisher information in Eq.~\eqref{eq:bounds} and Eq.~\eqref{eq:QCRB}}
In this section we evaluate the quantum Fisher information $H$ for the estimation of $\Delta\omega=\omega_2-\omega_1$ that is encoded in the input state $\ket{\bm{\psi}}$ defined in eq~\ref{app:input}. In order to do that, we evaluate first the quantum Fisher information for the single parameters $\omega_1$  and $\omega_2$. Each photon encode respectively one of these parameters. We define for the s-th photon the single-photon state $\ket{\psi_s}$, $s=1,2$, such that $\ket{\bm{\psi}}=\ket{\psi_1}\otimes\ket{\psi_2}$, where
\begin{align}
\begin{split}    \ket{\psi_1}=&\int_{\mathbb{R}}dt'_1\left\vert\psi\pto{t'_1}\right\vert \mathrm{e}^{-i\omega_1 t'_1}\opc{a}{1}(t'_1)\ket{0},\\
\ket{\psi_2}=&\int_{\mathbb{R}}dt'_2\left\vert\psi\pto{t'_2}\right\vert \mathrm{e}^{-i\omega_2 t'_2}\opc{b}{2}(t'_2)\ket{0}. \label{app:psii}
\end{split}
\end{align}
therefore, the quantum Fisher information is the sum of two matrices, i.e. $H=H_1+H_2$, where
\begin{equation}
H_s\pto{\omega_1, \omega_2}=4\mathrm{Re}\pto{\braket{\partial_{x}\psi_s\vert\partial_{x}\psi_s}-\left\vert\braket{\psi_s\vert\partial_{x}\psi_s}\right\vert^2}, \, x=\omega_1, \omega_2.\label{app:qfidef}
\end{equation}
Since each photon encode only one parameter, $\ket{\partial_{\omega_2}\psi_1}=\ket{\partial_{\omega_1}\psi_2}=0$. Therefore, the quantum Fisher information will be diagonal:
\begin{equation}
H\pto{\omega_1,\omega_2}= \begin{pmatrix}
H_1\pto{\omega_1}&0\\
0&H_2\pto{\omega_2}
\\ \end{pmatrix}.
\end{equation}
We find the s-th diagonal term by using $\ket{\psi_s}$ in Eq.~\eqref{app:psii} and $\ket{\partial_{\omega_s}\psi_s}$, as shown in Eq.~\eqref{app:qfidef}. The kets $\ket{\partial_{\omega_s}\psi_s}$ are the following:
\begin{align}
\begin{split}
\ket{\partial_{\omega_1}\psi_1}=&-i\int_{\mathbb{R}}dt'_1 t'_1\left\vert\psi\pto{t'_1}\right\vert \mathrm{e}^{-i\omega_1 t'_1}\opc{a}{1}(t'_1)\ket{0},\\
\ket{\partial_{\omega_2}\psi_2}=&-i\int_{\mathbb{R}}dt'_2 t'_2\left\vert\psi\pto{t'_2}\right\vert \mathrm{e}^{-i\omega_2 t'_2}\opc{b}{2}(t'_2)\ket{0}.\label{app:partialpsii}
\end{split}
\end{align}
So, by using the Eqs.~\eqref{app:psii},~\eqref{app:qfidef} and~\eqref{app:partialpsii}, we find the s-th diagonal term of the quantum Fisher information
\begin{equation}
H_s\pto{\omega_s}=4\pto{\int_{\mathbb{R}}dt_s\left\vert t_s\psi\pto{t_s}\right\vert^2-\left\vert\int_{\mathbb{R}}dt_s t_s\left\vert \psi_s\pto{t_s}\right\vert^2\right\vert^2}=4\tau^2,
\end{equation}
where $\tau$ is the variance of the temporal distribution $\left\vert\psi\pto{t}\right\vert^2$ of each photon. Another approach to obtain the quantum Fisher information uses the fact that the probe state is pure and that the evolution operator is unitary, with generator $\hat{T}_s$ represented in the temporal domain as $\hat{T}_s=i\frac{\partial}{\partial \omega_s}$ for the photon in the s-th input channel, such that $\ket{\psi_s}=\exp(-i\omega_s\hat{T}_s)\ket{\psi_{s,0}}$, where $\ket{\psi_{s,0}}$ is the initial state of the s-th photon at a frequency equal to zero (a similar and more detailed description for the evaluation of the quantum Fisher information using these kinds of operators can be found in~\citep{PhysRevA.106.063715}). With this notation, $\ket{\partial_{\omega_s}\psi_s}=-i\hat{T}_s\exp(-i\omega_s\hat{T}_s)\ket{\psi_{s,0}}=-i\hat{T}_s\ket{\psi_s}$, and therefore, using Eq.~\eqref{app:qfidef}, it is possible to find that
\begin{equation}
    H_s\pto{\omega_s}=4(\langle \hat{T^2_s}\rangle-\langle \hat{T_s}\rangle\langle \hat{T_s}\rangle)=4\mathrm{Var}[\hat{T}_s],
\end{equation}
where the expected values $\langle\star\rangle$ and the variance $\mathrm{Var[\star]}$ are evaluated using the state $\ket{\psi_{s}}$. Using Eq.~\eqref{app:psii} and recalling that the generator in the temporal domain can be represented as $\hat{T}_s=i\frac{\partial}{\partial \omega_s}$, we have
\begin{align}
    \begin{split}
         H_s\pto{\omega_s}&=4\left(-\bra{\psi_s}\frac{\partial^2}{\partial\omega_s^2}\ket{\psi_s}+\bra{\psi_s}\frac{\partial}{\partial\omega_s}\ket{\psi_s}\bra{\psi_s}\frac{\partial}{\partial\omega_s}\ket{\psi_s}\right)\\
         &=4\pto{\int_{\mathbb{R}}dt_s\left\vert t_s\psi\pto{t_s}\right\vert^2-\left\vert\int_{\mathbb{R}}dt_s t_s\left\vert \psi_s\pto{t_s}\right\vert^2\right\vert^2}=4\tau^2
    \end{split}
\end{align}

Thus, the quantum Fisher information for the parameters $\omega_1$  and $\omega_2$ assumes the form
\begin{equation}
H\pto{\omega_1,\omega_2}=4\tau^2 I,
\end{equation}
where $I$ is the $2\times2$ identity matrix. To evaluate the quantum Fisher information for $\Delta\omega$, we define $\omega_M=\omega_1+\omega_2$. We evaluate the quantum Fisher information for $\pto{\omega_M,\Delta\omega}$ by applying a transformation on $H$ with the Jacobian 
\begin{equation}
    J=\frac{1}{2}\begin{pmatrix}
        1&1\\
1&-1
    \\\end{pmatrix}
\end{equation}
in the following way:
\begin{align}
\begin{split}
H\pto{\omega_M,\Delta\omega}&=JH\pto{\omega_1,\omega_2}J^T\\&=\tau^2\begin{pmatrix}
1&1\\
1&-1
\\ \end{pmatrix}\begin{pmatrix} 
1&0\\
0&1
\\\end{pmatrix}\begin{pmatrix}
1&1\\
1&-1
\\\end{pmatrix}=2\tau^2I.
\end{split}
\end{align}
This quantum Fisher information proves that it is possible to estimate separately $\omega_M$ and $\Delta\omega$, and that the quantum Fisher information for $\Delta\omega$ is $H\pto{\Delta\omega}=2\tau^2$.

\setcounter{equation}{0}
\section{Evaluation of the Fisher information} \label{app:Fi}
\subsection{Fisher Information in Eq.~\eqref{eq:fi}}
In this section we evaluate the Fisher information starting from its definition
\begin{equation}
F_\nu\pto{\Delta\omega}=E\pq{\pto{\frac{\partial}{\partial\Delta\omega}\log  P_\nu\pto{\Delta t,X|\Delta\omega}}^2}.
\end{equation}
Here, by using the probabilities formula in Eq.~\eqref{eq:probs}, we find that
\begin{align}
\begin{split}
F_\nu\pto{\Delta\omega}
&=\sum_{X=A,B}\int_{\mathbb{R}}d\Delta t \frac{1}{P_\nu\pto{\Delta t,X|\Delta\omega}}\pto{\frac{\partial}{\partial\Delta\omega}P_\nu\pto{\Delta t,X|\Delta\omega}}^2\\
&=\nu^2\int_{\mathbb{R}}d\Delta t C\pto{\Delta t} \frac{\Delta t^2\sin^2\pto{\Delta\omega\Delta t}}{1-\nu^2\cos^2\pto{\Delta\omega\Delta t}}\\&=\int_{\mathbb{R}}d\Delta t C\pto{\Delta t}\Delta t^2\beta_\nu\pto{\Delta\omega\Delta t},\label{app:figeneral}
    \end{split}
\end{align}
where we define the function $\beta_\nu\pto{\Delta\omega\Delta t}$ as follows:
\begin{equation}
    \beta_\nu\pto{\Delta\omega\Delta t}=\frac{\nu^2\sin^2\pto{\Delta\omega\Delta t}}{1-\nu^2\cos^2\pto{\Delta\omega\Delta t}}.
\end{equation}
It is worth to notice that for $\nu=1$,  $\beta_\nu=1\pto{\Delta\omega\Delta t}=1$. So, in this case the Fisher information assumes the form
\begin{equation}
  F_{\nu=1}\pto{\Delta\omega}=  \int_{\mathbb{R}}d\Delta t \Delta t^2 C\pto{\Delta t}=2\tau^2=H\pto{\Delta\omega}.
\end{equation}
Therefore, for $\nu=1$, the fisher information saturates the quantum Cram\'{e}r-Rao bound, and the maximum precision is achieved.
\subsection{For $\tau\Delta\omega\gg 1$}
the Fisher information assumes a simpler form in the regime $\tau\Delta\omega\gg 1$. In fact, in this regime, we can approximate the value of the Fisher information by splitting the interval of integration in Eq.~\eqref{app:figeneral} in intervals of length $\pi/\Delta t$. In each of this intervals, the term $C\pto{\Delta t}\Delta t^2$ can be considered approximately constant, and thus we can substitute it with a constant value:
\begin{align}
\begin{split}
    F_\nu\pto{\Delta\omega}&\simeq \sum_{n\in \mathbb{Z}}\pto{\frac{\pi n}{\Delta\omega}}^2C\pto{\frac{\pi n}{\Delta\omega}}\int_{\frac{n\pi}{\Delta\omega}}^{\frac{\pto{n+1}\pi}{\Delta\omega}}d\Delta t \beta_\nu\pto{\Delta\omega\Delta t}\\&=\sum_{n\in \mathbb{Z}}\pto{\frac{\pi n}{\Delta\omega}}^2C\pto{\frac{\pi n}{\Delta\omega}}\frac{\Delta\omega}{\pi}\bar{\beta}_\nu,\label{app:fisum}
    \end{split}
\end{align}
where $\bar{\beta}_\nu$ is the mean of the function $\beta_\nu\pto{x}$ in each interval:
\begin{equation}
   \bar{\beta}_\nu=\frac{\Delta\omega}{\pi}\int_{0}^{\frac{\pi}{\Delta\omega}}d\Delta t \beta_\nu\pto{\Delta\omega\Delta t}=\frac{1}{\pi}\int_0^\pi dx\beta_\nu\pto{x}=1-\sqrt{1-\nu^2}.
\end{equation}
By summing again all the terms of the Fisher information in Eq.~\eqref{app:fisum}, we can rewrite the Fisher information as an integral, by using the approximations $\sum_{n\in \mathbb{Z}}\frac{\Delta\omega}{\pi}\bar{\beta}_\nu\rightarrow\int_\mathbb{R} d\xi$ and $\pto{\frac{\pi n}{\Delta\omega}}^2C\pto{\frac{\pi n}{\Delta\omega}}\rightarrow C\pto{\xi}\xi^2$,
\begin{equation}
F_\nu\pto{\Delta\omega}\simeq\bar{\beta}_\nu\int_\mathbb{R}d\xi C\pto{\xi}\xi^2=2\tau^2\bar{\beta}_\nu.
\end{equation}
In this way we prove that in the regime $\tau\Delta\omega\gg 1$, the Fisher information is proportional to the quantum Fisher information by a function of only $\nu$.

\subsection{Fisher information in Eq.~\eqref{eq:fig}}
If the distribution in time of the two input photons $\left\vert \psi\pto{t} \right\vert^2$ is Gaussian, we can show that the envelope in Eq.~\eqref{app:env} can be expressed as follows:
\begin{equation}
    C\pto{\Delta t}=\frac{1}{\sqrt{4\pi\tau^2}}\mathrm{exp}\pq{-\frac{\Delta t^2}{4\tau^2}}.\label{app:gaussenv}
\end{equation}
Therefore, by substituting $\Delta t/2\tau\rightarrow\kappa$ in Eq.~\eqref{app:figeneral}, we found that the Fisher information for Gaussian photons in time is
\begin{align}
\begin{split}
    &F^{G}_\nu\pto{\Delta\omega}=\frac{4\tau^2}{\sqrt{\pi}}\int_{\mathbb{R}}d\kappa \mathrm{e}^{-\kappa^2}\kappa^2\beta_\nu\pto{2\tau\Delta\omega\kappa}\\&\Rightarrow \frac{F_\nu\pto{\Delta\omega}}{H\pto{\Delta\omega}}=\frac{2}{\sqrt{\pi}}\int_{\mathbb{R}}d\kappa \mathrm{e}^{-\kappa^2}\kappa^2\beta_\nu\pto{2\tau\Delta\omega\kappa}.
\end{split}
  \end{align}  
\subsection{Fisher information in Eq.~\eqref{eq:fignr}}
If the detectors at the output of the beam splitter represented in \figurename~\ref{fig:setup} do not resolve the time delay, the output can be only of two types: bunching and coincidence. We find their probabilities by integrating with respect to $\Delta t$ the probabilities in Eq.~\eqref{app:probs}: 
\begin{align}
\begin{split}
    P\pto{X|\Delta\omega}&=\int_\mathbb{R} d\Delta t P_\nu\pto{X,\Delta t|\Delta\omega}\\&=\frac{1}{2}+\frac{\nu\alpha\pto{X}}{2}\int_\mathbb{R} d\Delta t C\pto{\Delta t}\cos^2\pto{\Delta t\Delta\omega}, \, X=A,B.
    \end{split}
\end{align}
In particular, if the photons are Gaussian in time, by using Eq.~\eqref{app:gaussenv}, we find that
\begin{equation}
    P\pto{X|\Delta\omega}=\frac{1}{2}\pto{1+\nu\alpha\pto{X}\mathrm{e}^{-\tau^2\Delta\omega^2}}, \, X=A,B.\label{app:probgnr}
\end{equation}
We evaluate the Fisher information using its definition, that here is written for clarity:
\begin{equation}
F^{G, NR}_\nu\pto{\Delta\omega}=E\pq{\pto{\frac{\partial}{\partial\Delta\omega}\log  P_\nu\pto{X|\Delta\omega}}^2}.
\end{equation}
Therefore, by using Eq.~\eqref{app:probgnr}, the Fisher information assumes the form
\begin{align}
\begin{split}
    F^{G, NR}_\nu\pto{\Delta\omega}&=\sum_{X=A,B}\frac{1}{P_\nu\pto{X|\Delta\omega}}\pto{\frac{\partial P_\nu\pto{X|\Delta\omega}}{\partial\Delta\omega}}^2\\
    &=\frac{4\nu^2\tau^4\Delta\omega^2}{\mathrm{e}^{2\tau^2\Delta\omega^2}-\nu^2}=H\pto{\Delta\omega}\frac{2\nu^2\tau^2\Delta\omega^2}{\mathrm{e}^{2\tau^2\Delta\omega^2}-\nu^2}.
    \end{split}
\end{align}
For $\tau\Delta\omega\ll1$ we can simplify it by substituting $\mathrm{e}^{2\tau^2\Delta\omega^2}\rightarrow1+2\tau^2\Delta\omega^2$. In this case, the Fisher information becomes equal to the Quantum Fisher information:
\begin{equation}
    F^{G, NR}_\nu=1\pto{\Delta\omega}=2\tau^2=H\pto{\Delta\omega}.
\end{equation}
\end{appendices}
\bibliography{main}
\end{document}